\documentclass[lettersize,journal]{IEEEtran}

\usepackage{blindtext}
\usepackage{multicol}
\usepackage[cmex10]{amsmath}
\usepackage{etoolbox}        

\usepackage{graphicx}        
\graphicspath{{figures/}}

\usepackage{rotating}
\usepackage{adjustbox}

\usepackage{tikz}
\usepackage{xcolor}
\usepackage{mathtools}

\DeclarePairedDelimiterX\Basics[1](){ #1}

\usepackage{dsfont}
\usepackage{amsmath,amssymb}

\usepackage{makecell}
\setcellgapes{1.5pt}
\usepackage{amsmath}
\usepackage{stackengine}

\usepackage{bm}
\usepackage{array}
\usepackage{tabu}
\usepackage{multirow}
\usepackage{graphicx}

\usepackage[colorlinks=true, linkcolor    = blue, urlcolor     = blue, citecolor=blue, pdfborder={0 0 0}]{hyperref}
\usepackage[linesnumbered,lined,boxed,commentsnumbered,ruled]{algorithm2e}
\usepackage{float}
\usepackage{subcaption}
\usepackage{bm}
\usepackage{graphicx}
\usepackage{amsmath}
\usepackage{soul}
\usepackage{multirow}
\usepackage{amsthm,amssymb,amsmath}
\usepackage{mathrsfs}
\DeclareMathAlphabet{\mathpzc}{OT1}{pzc}{m}{it}
\usepackage{mathtools}

\usepackage{stackengine}

\theoremstyle{remark}

\newtheorem{theorem}{Theorem}
\newtheorem{example}{Example}

\newtheorem{lemma}{Lemma}

\usepackage{fontawesome5}
\usepackage{academicons}
\usepackage{xcolor}
\usepackage{amsmath}
\definecolor{dukeblue}{rgb}{0.0, 0.0, 0.61}
\definecolor{harvardcrimson}{rgb}{0.79, 0.0, 0.09}

\newcommand{\txbl}[1]{\textcolor{black}{#1}}

\newcommand{\orcid}[1]{\href{https://orcid.org/#1}{\textcolor[HTML]{A6CE39}{\faOrcid}}}

\usepackage{amsmath}
\usepackage{cite}
\usepackage{xfrac}
\usepackage{graphicx}
\usepackage[font={small}]{caption}
\usepackage{ptext}
\usepackage{float}
\usepackage{setspace}
\DeclareSymbolFont{matha}{OML}{txmi}{m}{it}
\DeclareMathSymbol{\varv}{\mathord}{matha}{118}
\usepackage{amsmath}
\usepackage{graphicx}
\newcommand{\alabel}[1]{\stepcounter{equation}\tag{\theequation}\label{#1}}

\newcommand{\vect}[1]{\textbfit{#1}}

\newcommand{\vX}{\vect{X}}
\newcommand{\vx}{\vect{x}}

\newcommand{\vy}{\vect{y}}

\newcommand{\set}[1]{\mathcal{#1}}

\newcommand{\setS}{\set{S}}

\newcommand{\pI}{\pi_{\mathrm{I}}}
\newcommand{\pD}{\pi_{\mathrm{D}}}
\newcommand{\pS}{\pi_{\mathrm{S}}}

\newcommand{\defeq}{\triangleq}

\newcommand{\Z}{\mathbb{Z}}

\newcommand{\exampleend}{\mbox{}\hfill$\square$}

\newcounter{mytempeqcounter}

\newcommand{\bigformulatop}[2]{%
	\begin{figure*}[!t]
		\normalsize
		\setcounter{mytempeqcounter}{\value{equation}}
		\setcounter{equation}{#1}
		#2
		\setcounter{equation}{\value{mytempeqcounter}}
		\vspace*{-10pt}
		\hrulefill
		\vspace*{-6pt}
	\end{figure*}
}

\makeatletter
\DeclareRobustCommand\bfseriesitshape{%
	\not@math@alphabet\itshapebfseries\relax
	\fontseries\bfdefault
	\fontshape\itdefault
	\selectfont
}
\makeatother

\DeclareTextFontCommand{\textbfit}{\bfseriesitshape}

\begin{document}

\title{Belief-Combining Framework for Multi-Trace Reconstruction over Channels with Insertions, Deletions, and Substitutions}

\author{Aria~Nouri\textsuperscript{\orcid{0000-0001-5548-184X}}
		\thanks{\\\rule{\linewidth}{0.4pt}\\[4pt] This study was conducted independently; (correspondence: \href{mailto:ariya@ieee.org}{\txbl{ariya@ieee.org}}).\\
		\rule{\linewidth}{0.4pt}		
} 
}

\markboth{}%
{Nouri: Belief-Combining Framework for Multi-Trace Reconstruction over Channels with Insertions, Deletions, and Substitutions}

\maketitle

\begin{abstract}
Optimal reconstruction of a source sequence from multiple noisy traces corrupted by random insertions, deletions, and substitutions typically requires joint processing of all traces, leading to computational complexity that grows exponentially with the number of traces. In this work, we propose an iterative belief-combining procedure that computes symbol-wise a posteriori probabilities by propagating trace-wise inferences via message passing. We prove that, upon convergence, our method achieves the same reconstruction performance as joint maximum a posteriori estimation, while reducing the complexity to quadratic in the number of traces. This performance equivalence is validated using a real-world dataset of clustered short-strand DNA reads.
\end{abstract}


\section{Introduction}
\IEEEPARstart{S}{ynthetic} DNA has emerged as a promising medium for data storage, attracting significant interest from coding and information theory communities in the pursuit of reliable data retention~\cite{10440314}. Each information sequence written into a DNA storage system generates multiple DNA strands during synthesis and polymerase chain reaction amplification. Upon retrieval, these strands must first be clustered according to their unknown originating sequences. Within each cluster, the original sequence is reconstructed from multiple noisy reads that may contain random insertion, deletion, and substitution (IDS) errors, giving rise to the trace reconstruction problem~\cite{US2022166446A1,9889680,10889909,9174050,US20180211001A1}.




Among the classic works on maximum a posteriori (MAP) estimation over a single IDS channel, two Markov-based approaches continue to draw attention in the context of reconstructing a coded sequence from multiple noisy traces. The first is the work of Bahl and Jelinek~\cite{1055419}, later generalized to the coded multi-trace setting in~\cite{US2022166446A1}. In this approach, the hidden Markov state of the IDS channel is a variable that tracks the \emph{output pointer}, i.e., the position of the output symbol corresponding to the on-deck input symbol. The second is the work of Davey and MacKay~\cite{910582}, later extended to the coded multi-trace setting in~\cite{9889680}. Here, the hidden Markov state, referred to as the \emph{drift}, is defined as the cumulative number of insertions minus deletions prior to the on-deck input symbol. This drift variable can also be interpreted as tracking the output pointer as a function of the input position similar to the first approach, making the two approaches conceptually equivalent.

Both works~\cite{US2022166446A1,9889680} consider the same two general frameworks for MAP estimation in the multi-trace setting. The first framework constructs a joint Markov model that incorporates all received traces. The BCJR algorithm~\cite{1055186} is then applied~to the resulting multi-trace trellis, yielding optimal symbol-wise a posteriori probabilities (APPs). However, the number of joint-trellis edges grows exponentially with the number of traces, and since the complexity of BCJR scales linearly with the number of edges, the overall decoding complexity becomes exponential in the number of traces. To alleviate this complexity, both works propose a second framework in which each trace is decoded independently on its own local trellis and the resulting estimates are subsequently combined; while this approach reduces the complexity to linear in the number of traces, it~incurs a substantial loss in their reconstruction~performance.


The authors of~\cite{US2022166446A1} proposed a method to improve reconstruction performance over separate trellises by injecting the hard estimate from the decoder of each individual trace into the decoder of the subsequent trace. This approach outperforms fully independent decoding followed by post-combination. However, as confirmed by simulations for two and three traces, a noticeable performance gap remains relative to joint-trellis decoding, and no results are reported beyond this range due to the prohibitive complexity of reconstruction over joint trellises.

In this work, we propose a belief-propagation–based framework that processes each trace on its own local trellis. The soft beliefs produced during the decoding of each trace are iteratively exchanged via message passing with the decoders of adjacent traces according to a prescribed ordering. We show that the proposed framework achieves MAP performance equivalent to that of BCJR applied to the joint trellis, while reducing the decoding complexity to quadratic growth in the number of traces. To the best of our knowledge, this is the first approach to attain joint-trellis MAP performance with a decoding complexity that does not grow exponentially in the number of traces~\cite{10889909}. Existing approaches that attempt to exploit joint information across traces either incur exponential complexity or rely on suboptimal or heuristic decoding strategies~\cite{9174050,US2022166446A1}.

The proposed framework is general; as long as the trellis associated with a Markov model forms a directed-acyclic graph, our proposed approach can combine beliefs from separate trellises regardless of the nature of the hidden state. For instance, the channel state can be merged with an encoder state to form a joint state variable, and our approach can then be applied to estimate this joint state. This makes the framework readily applicable in the coded setting. Moreover, since the output is in the form of soft APPs, the method can be naturally integrated into concatenated coding schemes with outer codes.

The rest of this letter is organized as follows:
Section~\ref{sec:prelim} describes preliminary concepts;
Section~\ref{sec:bc} presents the central contributions;
Section~\ref{sec:results} reports simulation results;
finally, Section~\ref{sec:conclusion} concludes the letter and outlines future directions.

Throughout the paper, random variables are denoted by upper-case italic letters (e.g., $X$) with realizations in lower-case (e.g., $x$). Random vectors are denoted by upper-case boldface italic letters (e.g., $\vX$) with realizations in lower-case (e.g., $\vx$). For integers $n_1 \le n_2$, we write $\vect{X}_{[n_1:n_2]} \defeq (X_{n_1},\ldots,X_{n_2})$ and $\vect{x}_{[n_1:n_2]} \defeq (x_{n_1},\ldots,x_{n_2})$. The sets of integers and non-negative integers are respectively denoted by $\Z$ and $\mathbb{Z}_{\ge 0}$, the floor operator by $\lfloor \cdot \rfloor$, and the probability of an event by $\Pr(\cdot)$.

\section{Preliminaries}\label{sec:prelim}

\begin{figure}
	\centering
	\includegraphics[scale=0.82]{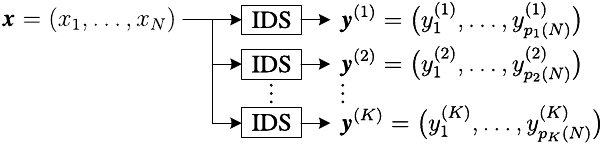}
	\caption{Multi-trace insertion–deletion–substitution (IDS) channel.\vspace{4pt}}\label{fig:IDS}
\end{figure}

Let $\vect{x} \defeq (x_1,\ldots,x_N)$ denote the information sequence to be transmitted over multiple IDS channels (Fig.~\ref{fig:IDS}), where $x_t \in \Sigma$ for $1 \le t \le N$ and $\Sigma$ is a finite alphabet. The IDS channels produce $K$ noisy output traces, denoted by $\vect{y}^{(k)}$, $1 \le k \le K.$

Due to stochastic insertion and deletion events, the output indices associated with the input sequence are modeled as a random process $\{P(t)\}_{t=1}^N$, where $P(t) \in \mathbb{Z}_{\ge 0}$ denotes the index of the output symbol $Y_{P(t)}\in\Sigma$ corresponding to the input symbol $X_t$. Each received trace provides an independent realization of this process, denoted by $\{p_k(t)\}_{t=1}^N$ for the $k$-th trace. Accordingly, whenever $p_k(t) \ge 1$, the output symbol associated with $X_t$ at the $k$-th trace is denoted by $y^{(k)}_{p_k(t)} \in \Sigma$.

For an on-deck input symbol $x_t \in \Sigma$: 
(i) with probability~$\pI$, the IDS channel inserts a symbol drawn uniformly at random from $\Sigma$ without consuming $x_t$, advancing the output index as $p_k(t) \rightarrow p_k(t)+1$; 
(ii) with probability $\pD$, $x_t$ is deleted, no output symbol is produced, and the channel queues the next input, $p_k(t) \rightarrow p_k(t+1)$; 
(iii) with probability $\pS$, $x_t$ is substituted by a symbol drawn uniformly at random from $\Sigma \setminus \{x_t\}$, the next input is queued, and the output index advances as $p_k(t) \rightarrow p_k(t+1)+1$; 
(iv) with probability $1-\pI-\pD-\pS$, $x_t$ is transmitted without error, the channel queues the next input, and the output index advances as $p_k(t) \rightarrow p_k(t+1)+1$.

Let $S_i^{(k)} \in \mathcal{S}$ denote the Markov state of the channel experienced by the $k$-th trace at trellis stage $i$, where $\mathcal{S}$ is a finite state alphabet and $1 \le k \le K$.\footnote{One of the parameters encoded by the trellis state is the output pointer $p_k(\cdot)$. Since different traces may have different lengths, the terminal output position $p_k(N)$ can vary across $k$, potentially leading to different state-alphabet cardinalities  for the corresponding Markov models. Nevertheless, assuming that the output pointer does not drift by more than $\Delta < N$ symbols from the input index, the state alphabet $\setS$ can be taken to be identical across all received traces. Consequently, the trace index $k$ is omitted from the state alphabet, except in Example~\ref{ex:trellisBMA}, where the full trellis is shown for illustration.} Let $q(i) \in \mathbb{Z}$ denote the index of the input symbol $X_{q(i)}$ associated with trellis stage $i$. Since the composite mapping $p_k(q(\cdot))$ appears frequently, we introduce the shorthand $\zeta_k(\cdot) \defeq p_k\bigl(q(\cdot)\bigr)$, which specifies the index of the output symbol in the $k$-th trace corresponding to a given trellis stage. Accordingly, the output symbol emitted on a transition terminating at state $S_i^{(k)}$ is denoted by $Y^{(k)}_{\zeta_k(i)}$.

A trellis branch is specified by $B_{i^-,i}^{(k)} \defeq \bigl(S_{i^-}^{(k)},\, Y^{(k)}_{\zeta_k(i)},\, S_i^{(k)}\bigr)$, representing a transition from a predecessor state $S_{i^-}^{(k)}\in\setS$ to a successor state $S_i^{(k)}\in\setS$ that emits the output symbol $Y^{(k)}_{\zeta_k(i)}$. The predecessor index $i^-$ may correspond to the same trellis stage as $i$ ($i^- = i$) or to an earlier stage ($i^- < i$), consistent with the directed-acyclic structure of the trellis. If a transition produces no output symbol, as in a deletion event, the corresponding branch is written as $B_{i^-,i}^{(k)}\, \defeq\, \bigl(S_{i^-}^{(k)},\, \Phi,\, S_i^{(k)}\bigr).$

\begin{example}\label{ex:trellisBMA}
	For illustration, consider the trellis representation associated with the $k$-th trace of a pointer-based IDS channel Markov model~\cite{1055419, US2022166446A1}, shown in Fig.~\ref{fig:Trellis-BMA-section}. In this setting, each trellis section comprises four consecutive states $S_{4(t-1)}^{(k)}, \ldots, S_{4t-1}^{(k)}$, which collectively describe the evolution of the output pointer $\zeta_k(i)$ over $4(t-1) \le i \le 4t-1$ while processing the on-deck input index $t$. The mapping from trellis stage index $i$ to input index $t$ is given by $q(i) = \lfloor i/4 \rfloor + 1$, so that all states within the $t$-th trellis section satisfy $q(i) = t$, uniquely identifying the associated on-deck input symbol $X_t$.\footnote{The trellis section is identical for all $t \in \mathbb{Z}$, so the trellis is periodic with respect to $t$, and a single section suffices to illustrate all valid state transitions.}
	
	The $t$-th trellis section begins at the initial state $s_{4t-4}^{(k)} \defeq p_k(t)$. 	The outgoing branches from this state prepare the buffer for processing $X_t$ across all admissible output positions $1 \le \zeta_k(i) \le p_k(N)$. The subsequent two intermediate states are given by $s_{4t-3}^{(k)} \defeq \bigl(p_k(t), x_t\bigr)$ and $s_{4t-2}^{(k)} \defeq \bigl(p_k(t+1), x_t\bigr)$; the overall state space of the trellis therefore has cardinality $|\mathcal{S}^{(k)}| = (p_k(N)+1)\,|\Sigma|$. The corresponding transitions propagate the symbol-wise APPs of $X_t$ jointly with the associated output position $P(t)$ across the trellis. The trellis section concludes by flushing the buffer corresponding to $X_t$ and terminating at state $s_{4t-1}^{(k)} \defeq p_k(t+1)$, thereby preparing the input buffer for processing the subsequent input symbol $X_{t+1}$.

	\begin{figure}[t]
		\centering
		\includegraphics[scale=0.686]{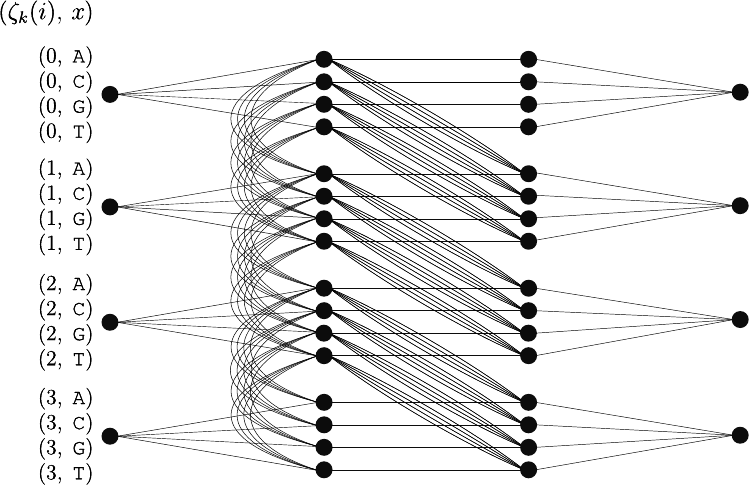}
		\caption{Trellis section of the pointer-based IDS channel Markov model for a received trace of length $p_k(N)=3,$ and $\Sigma=\{\texttt{A,C,G,T}\}$.\vspace*{6pt}}
		\label{fig:Trellis-BMA-section}
	\end{figure}
	
	The permissible transitions within the trellis are as follows: an insertion event results in $p_k(t) \to p_k(t)+1$ (vertical branches), a deletion event results in $p_k(t) \to p_k(t+1)$ (horizontal branches), and a substitution or noiseless transmission event results in $p_k(t) \to p_k(t+1)+1$ (diagonal branches). In this example, the four \emph{parallel vertical} branches correspond to the insertion of one symbol from $\Sigma$, while the four \emph{parallel diagonal} branches comprise a single noiseless transmission branch and $|\Sigma|-1 = 3$ substitution branches, collectively accounting for all possible symbol-level state transitions. 
	\exampleend
\end{example}

\section{Refined State Posteriors}\label{sec:bc}
Optimal MAP estimation in the multi-trace setting requires evaluating and making a decision based on the joint state APP
\begin{equation}
	\Pr\left(\left[S_i^{(1)}, \ldots, S_i^{(K)}\right] \middle|\, \vy^{(1)}, \ldots, \vy^{(K)}\right)\!,
	\label{eq:joint-state-posterior}
\end{equation}
where $\big[S_i^{(1)}\!\! \ldots S_i^{(K)}\big]\defeq\big(\zeta_1(i)\ldots\zeta_K(i), X_{q(i)}\big),$~\cite{US2022166446A1}. However, calculating \eqref{eq:joint-state-posterior} is generally intractable under \emph{local} trace-wise processing. Our central idea is that, rather than approximating~\eqref{eq:joint-state-posterior} directly, we compute refined per-trace state marginals
\begin{equation}
	\Pr\bigl(	S_i^{(k)} \bigm| \vy^{(1)}, \ldots, \vy^{(K)}\bigr),
	\qquad 1 \le k \le K,
	\label{equ:per-trace-marginal}
\end{equation}
by iteratively exchanging extrinsic local APPs across traces. In the following section, we will show that, under a mild consensus condition on the symbol-level posteriors, these per-trace marginals constitute sufficient statistics for recovering the input symbol MAP estimates implied by the joint APP in~\eqref{eq:joint-state-posterior}.

\subsection{Sufficient Statistics for Joint MAP Estimation}
\label{sec:joint-to-marginal}

Let $\{\zeta_k(i)\}$ denote a shorthand for the alphabet set $\{\zeta_k(i):1\leq\zeta_k(i)\leq p_k(N)\}$, used in summations to indicate that $\zeta_k(i)$ is summed over all values in its alphabet. The symbol-level input posterior $\Pr(X_{q(i)} = x |\, \vy^{(1)}, \ldots, \vy^{(K)})$, inferred from the refined per-trace marginal in \eqref{equ:per-trace-marginal} at the $k$-th trellis, is
\begin{equation}\label{equ:marginals}
	\Lambda_i^{(k)}(x) \defeq\!\! \sum_{\{\zeta_k(i)\}} \Pr\Bigl(S_i^{(k)} = \big(\zeta_k(i), x\big) \Bigm| \vy^{(1)}, \ldots, \vy^{(K)}\Bigr).
\end{equation}
The input posterior inferred from the joint state APP in~\eqref{eq:joint-state-posterior} is
\begin{align*}
	\Lambda_i^\ast(x) \defeq \hspace{-4pt}\sum_{\{\zeta_1(i)\}}\hspace{-4pt}\cdots\hspace{-4pt}\sum_{\{\zeta_K(i)\}}\hspace{-4pt} \Pr\left(\left[S_i^{(1)}, \ldots, S_i^{(K)}\right] \middle|\, \vy^{(1)}, \ldots, \vy^{(K)}\right)\!.
\end{align*}

\bigformulatop{3}{
	\begin{align}\label{equ:top}
		\Pr\Bigl(S_i^{(k)} = \big(\zeta_k(i), x\big) \Bigm| \vy^{(1)}, \ldots, \vy^{(K)}\Bigr)
		=\hspace{-12pt}\sum_{\{\zeta_{1}(i)\}\cdots\{\zeta_K(i)\}\setminus\{\zeta_k(i)\}}\hspace{-12pt} \Pr\bigg(\left[S_i^{(1)}, \ldots, S_i^{(K)}\right] = \Big(\zeta_1(i)\ldots\zeta_K(i),x\Big) \bigg|\, \vy^{(1)}, \ldots, \vy^{(K)}\bigg).
	\end{align}
}

\begin{figure}
	\centering
	\includegraphics[scale=0.425]{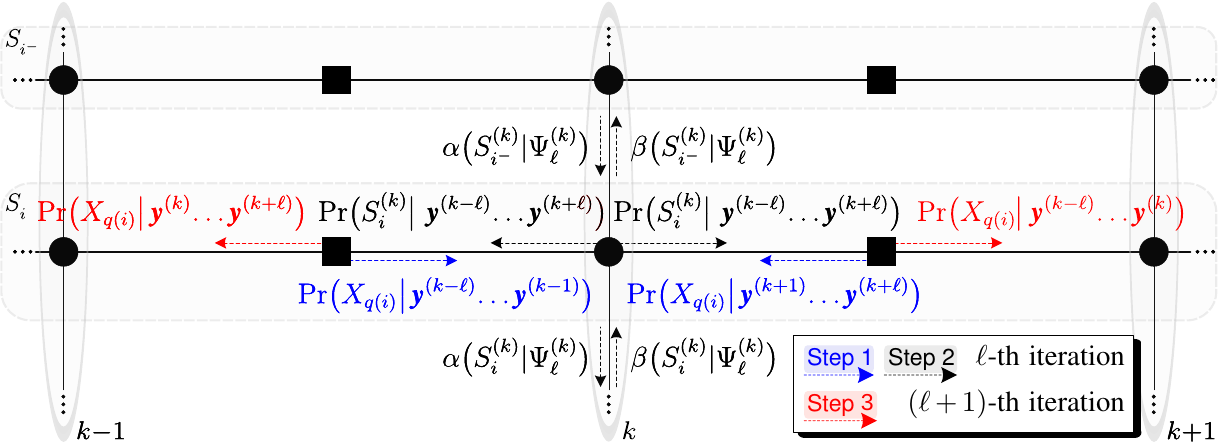}
	\caption{Message-passing updates across the simplified factor graph.\vspace{5pt}}\label{fig:BP}
\end{figure}

\begin{lemma}[\emph{Symbol-Level Consensus}]\label{lem}
	Upon convergence of message passing in the factor graph representing the trace-wise inference problem (Fig.~\ref{fig:BP}), all per-trace input APPs~\eqref{equ:marginals}, $\forall k,$ reach consensus on a common value $\Lambda_i^{(k)}(X_{q(i)})=\Lambda_i(X_{q(i)})$.
\end{lemma}

\begin{IEEEproof}
	Since the input symbol $X_{q(i)}$ is shared across all $K$ trellises, the factor graph representing the inference problem naturally enforces consistency constraints on the symbol-level marginals. Hence, any stationary point of the Bethe free energy for this graph satisfies marginal consistency across all factors associated with $X_{q(i)}$,~\cite{1459044}, (i.e., the horizontal factors in Fig.~\ref{fig:BP}). Consequently, upon convergence of message passing, all per-trace input APPs in~\eqref{equ:marginals} converge to a common value $\Lambda_i^{(k)}(X_{q(i)}) = \Lambda_i(X_{q(i)})$, for all $1 \le k \le K$. This \emph{symbol-level consensus} is consistently observed in simulations.
\end{IEEEproof}

\begin{theorem}
	The APP $\Lambda_i(X_{q(i)})$, inferred from each of the refined per-trace posteriors in \eqref{equ:marginals}, $\forall k$, is identical to $\Lambda_i^\ast(X_{q(i)}).$
\end{theorem}

\begin{IEEEproof}
	For all $1 \le k \le K$, we have from Lemma~\ref{lem} that
	\begin{align*}
		&\Lambda_i(x) = \Lambda_i^{(k)}(x) = \hspace{-2pt} \sum_{\{\zeta_k(i)\}} \hspace{-2pt} \Pr\Bigl(S_i^{(k)}\! = \big(\zeta_k(i), x\big) \Bigm| \vy^{(1)} \!\!\ldots \vy^{(K)}\Bigr)\\
		&\overset{(a)}{=}\hspace{-6.6pt}\sum_{\{\zeta_{1}(i)\}\cdots\{\zeta_K(i)\}}\hspace{-6.6pt} \Pr\left(\left[S_i^{(1)}, \ldots, S_i^{(K)}\right] \middle|\, \vy^{(1)} \!\!\ldots \vy^{(K)}\right) = \Lambda_i^\ast(x),
	\end{align*}
	where $(a)$ follows from \eqref{equ:top} at the top of the next page and the marginal consistency across $S_i^{(1)}, \ldots, S_i^{(K)}$, which is a direct consequence of the graphical model structure (Lemma~\ref{lem}).
\end{IEEEproof}

\subsection{Calculation of Refined Per-Trace Posteriors}
Following the discussions in the previous section, this section focuses on calculating per-trace state posteriors in~\eqref{equ:per-trace-marginal}.

The conventional BCJR algorithm~\cite{1055186} is first applied to all $K$ trellises independently, without any extrinsic prior, yielding $\Pr\big(S_i^{(k)}, \vect{y}^{(k)}\big),$ for $1 \le k \le K$. Normalizing over $\mathcal{S}$ gives us
\begin{align}\setcounter{equation}{4}
	\Pr\big(S_i^{(k)} | \vect{y}^{(k)}\big) &= 
	\frac{\Pr\big(S_i^{(k)}, \vect{y}^{(k)}\big)}
	{\sum_{s_i^{(k)} \in \mathcal{S}} \Pr\big(s_i^{(k)}, \vect{y}^{(k)}\big)},\>\> 1 \le k \le K.
	\label{equ:norm2}
\end{align}
The input APP produced by the $k$-th trellis is calculated by
$
	\Pr\bigl(X_{q(i)} = x \bigm|\vy^{(k)}\bigr) = \sum_{\{\zeta_k(i)\}} \Pr\bigl(S_i^{(k)}\!\! =  (\zeta_k(i), x) \bigm|\vy^{(k)}\bigr).
$
This APP is passed as extrinsic information to the neighboring trellises $k-1$ and $k+1$, marking the zeroth iteration ($\ell=0$).

At the $\ell$-th iteration ($\ell\geq 1$), our goal is to calculate
\begin{align}
	\Pr\!\big(S_i^{(k)} \,\big|\, \vect{y}^{(k-\ell)},\ldots, \vect{y}^{(k+\ell)}\big).
	\label{equ:norm2:comp}
\end{align}
Let $\Psi_\ell^{(k)}\defeq (\vect{y}^{(k-\ell)},\ldots, \vect{y}^{(k-1)}, \vect{y}^{(k+1)},\ldots, \vect{y}^{(k+\ell)})$. We begin by calculating $\Pr(S_i^{(k)}, \vect{y}^{(k)} | \Psi_\ell^{(k)}).$ This term is then normalized over alphabet $\mathcal{S}$ in the same manner as \eqref{equ:norm2}, to obtain~\eqref{equ:norm2:comp}.

Following the notation in the original BCJR paper~\cite{1055186}, we define the forward and backward state metrics as follows
\begin{align*}
	\alpha(S_i^{(k)} | \Psi_\ell^{(k)})
	&\defeq 
	\Pr\!\left(S_i^{(k)},\, \vect{y}^{(k)}_{[1:\zeta_k(i)]} \,\middle|\, 
	\Psi_\ell^{(k)} \right)\!, \alabel{equ:alpha} \\
	\beta(S_i^{(k)} | \Psi_\ell^{(k)})
	&\defeq 
	\Pr\!\left(\vect{y}^{(k)}_{[\zeta_k(i)+1:p_k(N)]} \,\middle|\, 
	S_i^{(k)}, \Psi_\ell^{(k)} \right)\!. \alabel{equ:beta}
\end{align*}
Also, the branch metric to be employed in the recursive calculation of \eqref{equ:alpha} and \eqref{equ:beta} is
\begin{align*}
	\gamma(B_{i^-,i}^{(k)} | \Psi_\ell^{(k)})
	\defeq \Pr\!\left(S_i^{(k)},\, {y}^{(k)}_{\zeta_k(i)}  \,\middle|\, S_{i^-}^{(k)}, \Psi_\ell^{(k)} \right)\!.\alabel{equ:gamma}
\end{align*}
Noting that $\zeta_k(i)\!=\!\zeta_k(i^-)+1$ for transitions emitting an output symbol and $\zeta_k(i)=\zeta_k(i^-)$ for deletions and other transitions that do not produce an output symbol, it follows directly that
\begin{align*}
	\Pr(S_i^{(k)}, \vect{y}^{(k)} | \Psi_\ell^{(k)})
	= \alpha(S_{i}^{(k)} | \Psi_\ell^{(k)})
	\cdot \beta(S_i^{(k)} | \Psi_\ell^{(k)}). \alabel{equ:rule}
\end{align*}

In the following, we discuss how the terms in \eqref{equ:alpha}, \eqref{equ:beta}, and \eqref{equ:gamma} can be calculated recursively. Prior to this, we outline how the incoming extrinsic beliefs from the neighboring traces $k-1$ and $k+1$ are combined to form a prior belief at the $k$-th trellis.

Given the APP $\Pr(X_{q(i)} \mid \vect{y}^{(k-\ell)},\ldots, \vect{y}^{(k-1)})$ passed from the $(k-1)$-th trellis and $\Pr(X_{q(i)} \mid \vect{y}^{(k+1)},\ldots, \vect{y}^{(k+\ell)})$ from the $(k+1)$-th trellis, the prior for the $k$-th trellis at iteration $\ell$ is constructed by combining these extrinsic beliefs as follows
\begin{align*}
	&\Pr\!\big(X_{q(i)} \,\big|\, \Psi_\ell^{(k)}\big)\approx\alabel{equ:BPprior}\\
	&
	\frac{
		\Pr\!\left(X_{q(i)} \middle| \vect{y}^{(k-\ell)},\ldots, \vect{y}^{(k-1)}\right)
		\cdot
		\Pr\!\left(X_{q(i)} \middle| \vect{y}^{(k+1)},\ldots,\vect{y}^{(k+\ell)}\right)
	}{
		\mathcal{Z}\cdot\Pr\!\left(X_{q(i)}\right)
	},
\end{align*}
where $\mathcal{Z}$ is a normalization factor. This is an approximation because the conditional independence $(\vect{y}^{(k-\ell)},\ldots, \vect{y}^{(k-1)}) \perp (\vect{y}^{(k+1)},\ldots,\vect{y}^{(k+\ell)}) \;\big|\; X_{q(i)}$ does not strictly hold as the traces remain coupled through the full input sequence $\vect{X}$. 
Nevertheless, multiplying the incoming beliefs is a standard extrinsic-message update rule in belief propagation. The resulting approximation errors from symbol-level dependencies are progressively reduced as iterative message passing continues, making the fused priors increasingly accurate over iterations.

We factorize the branch metric in~\eqref{equ:gamma} as follows
\begin{align*}
	\gamma(B_{i^-,i}^{(k)} | \Psi_\ell^{(k)})
	= &\Pr\!\big(S_i^{(k)} | S_{i^-}^{(k)}, \Psi_\ell^{(k)}\big)\\
	\cdot 
	&\Pr\left({y}^{(k)}_{\zeta_k(i)} \middle| S_{i^-}^{(k)}, S_i^{(k)}, \Psi_\ell^{(k)}\right)\!.\alabel{equ:gamma:factor}
\end{align*}
For trellis branches that import an input symbol into the buffer (see footnote~\ref{fn:branch} for other cases), the first factor of~\eqref{equ:gamma:factor} becomes
\begin{align*}
	&\Pr\big(S_i^{(k)} | S_{i^-}^{(k)},\,\Psi_\ell^{(k)}\big)
	=\Pr\big(\zeta_k(i),\,X_{q(i)} | \zeta_k(i^-),\,\Psi_\ell^{(k)}\big)\alabel{equ:resprior}\\
	&= \Pr\big(\zeta_k(i) | \zeta_k(i^-),\,\Psi_\ell^{(k)}\big)
	\cdot\Pr\big(X_{q(i)} | \zeta_k(i),\,\zeta_k(i^-),\,\Psi_\ell^{(k)}\big).
\end{align*}
By the independence of IDS events across traces, the pointer transition in the $k$-th trace is independent of the other traces, so
$ 
	\Pr\big(\zeta_k(i) | \zeta_k(i^-),\,\Psi_\ell^{(k)}\big)
	= \Pr\big(\zeta_k(i)| \zeta_k(i^-)\big),
$ 
where the right-hand side (RHS) is determined by the IDS event probabilities. Moreover, when the $k$-th trace $\vect{y}^{(k)}$ is not observed, the posterior of the input symbol $X_{q(i)}$ does not depend on the pointer indices of the $k$-th trace, and therefore
$ 
	\Pr\big(X_{q(i)} | \zeta_k(i),\,\zeta_k(i^-),\,\Psi_\ell^{(k)}\big)
	= \Pr\big(X_{q(i)} | \Psi_\ell^{(k)}\big),
$ 
where the RHS is available from \eqref{equ:BPprior}.\footnote{The extrinsic priors are applied only to branches that import an input symbol into the buffer (e.g., the first stage of Fig.~\ref{fig:Trellis-BMA-section}). For branches modeling IDS events (the intermediate stage of Fig.~\ref{fig:Trellis-BMA-section}), \eqref{equ:resprior} reduces to the pointer-transition term because the input symbol remains unchanged; $\Pr(S_i^{(k)} | S_{i^-}^{(k)}, \Psi_\ell^{(k)}) = \Pr(\zeta_k(i), X_t | \zeta_k(i^-), X_t, \Psi_\ell^{(k)}) = \Pr(\zeta_k(i) | \zeta_k(i^-), \Psi_\ell^{(k)})$, where the last equality follows since $\zeta_k(i)$ and $X_t$ are independent in the absence of $\vect{y}^{(k)}$.\label{fn:branch}} Combining the above, the first factor on the RHS of \eqref{equ:gamma:factor} becomes
$ 
	\Pr\big(S_i^{(k)} | S_{i^-}^{(k)},\,\Psi_\ell^{(k)}\big)
	= \Pr\big(\zeta_k(i)| \zeta_k(i^-)\big)\cdot \Pr\big(X_{q(i)} | \Psi_\ell^{(k)}\big).
$ 
Furthermore, since ${y}^{(k)}_{\zeta_k(i)}$ is connected to the other traces only through the input sequence $\vect{X}$, and the information from the other trellises has already been incorporated into the priors of $X_{q(i)}$ in the first factor of \eqref{equ:gamma:factor} in \eqref{equ:resprior}, we approximate the second factor as
$ 
	\Pr({y}^{(k)}_{\zeta_k(i)} | S_{i^-}^{(k)},S_i^{(k)},\Psi_\ell^{(k)})
	\approx
	\Pr({y}^{(k)}_{\zeta_k(i)} | S_{i^-}^{(k)},S_i^{(k)}),
$ 
where the RHS represents the branch weight associated with the transition $(S_{i^-}^{(k)} \to S_i^{(k)})$ that emits the output symbol ${y}^{(k)}_{\zeta_k(i)}$. 
The approximation becomes exact when there is a single branch between the state pair $\big(S_{i^-}^{(k)}, S_i^{(k)}\big)$, or the branch produces no output, i.e.,
$ 
	\Pr\!\big(y^{(k)}_{\zeta_k(i)} | S_{i^-}^{(k)}\!, S_i^{(k)}\!, \Psi_\ell^{(k)}\big)
	\!=\! \Pr\!\big(\Phi | S_{i^-}^{(k)}\!, S_i^{(k)}\big)\! =\! 1.
$ 

By fixing the boundary condition $\alpha(s_0^{(k)} | \Psi_\ell^{(k)})$ for all $s_0^{(k)} \in \mathcal{S}$, the forward values in \eqref{equ:alpha} are calculated recursively for $i = 1, 2, \ldots$, proceeding sequentially through the directed trellis
\begin{align*}
	\alpha(S_i^{(k)} | \Psi_\ell^{(k)})
	= \sum_{s_{i^-}^{(k)} \in \mathcal{S}}
	\alpha(s_{i^-}^{(k)} | \Psi_\ell^{(k)})
	\cdot 
	\gamma(B_{i^-,i}^{(k)} | \Psi_\ell^{(k)}).\alabel{equ:forward}
\end{align*}

By fixing the boundary condition $\beta(s_{q^{-1}(N)}^{(k)} | \Psi_\ell^{(k)})$ for all $s_{q^{-1}(N)}^{(k)} \in \mathcal{S}$, the backward values in \eqref{equ:beta} are calculated recursively for $i = q^{-1}(N)-1, q^{-1}(N)-2, \ldots$, progressing decrementally from the final stage of the trellis
\begin{align*}
	\beta(S_{i^-}^{(k)} | \Psi_\ell^{(k)})
	= \sum_{s_{i}^{(k)} \in \mathcal{S}}
	\gamma(B_{i^-,i}^{(k)} | \Psi_\ell^{(k)})
	\cdot 
	\beta(s_{i}^{(k)} | \Psi_\ell^{(k)}).\alabel{equ:backward}
\end{align*}

Now we have all quantities in \eqref{equ:alpha} and \eqref{equ:beta} required to evaluate 
$\Pr(S_i^{(k)}, \vect{y}^{(k)} | \Psi_\ell^{(k)})$ via \eqref{equ:rule}. Next, we can calculate
\begin{equation}\label{equ:finalapp}
	\!\Pr\!\big(S_i^{(k)} | \vect{y}^{(k-\ell)}\!\!\!\!\ldots \vect{y}^{(k+\ell)}\big)
	\!=\!
	\frac{
		\Pr\big(S_i^{(k)}, \vect{y}^{(k)}
		| \Psi_\ell^{(k)}\big)
	}{
		\sum_{s_i^{(k)}\in\mathcal{S}}
		\Pr\!\big(s_i^{(k)}\!\!, \vect{y}^{(k)}
		| \Psi_\ell^{(k)}\big)
	}.\!
\end{equation}
The belief of the $k$-th trellis about the input symbol $X_{q(i)}$ at the $\ell$-th iteration is then calculated by
\begin{align*}
	&\Pr\bigl(X_{q(i)} = x \bigm|\vect{y}^{(k-\ell)},\ldots, \vect{y}^{(k+\ell)}\bigr)\\
	 &\quad= \sum_{\{\zeta_k(i)\}} \Pr\Bigl(S_i^{(k)} \! = \big(\zeta_k(i), x\big) \!\Bigm|\vect{y}^{(k-\ell)},\ldots, \vect{y}^{(k+\ell)}\Bigr).\alabel{equ:marginals:FB}
\end{align*}

Ultimately, the extrinsic belief passed from the $k$-th trellis to the $(k-1)$-th trellis at the $(\ell+1)$-th iteration is given by
\begin{align*}
	\Pr\!\big(X_{q(i)} \big|\, \vect{y}^{(k)}\!\ldots \vect{y}^{(k+\ell)}\big)\approx\frac{
		\Pr\!\big(X_{q(i)} \big|\, \vect{y}^{(k-\ell)}\!\ldots \vect{y}^{(k+\ell)}\big)
	}{
		\mathcal{Z} \cdot \Pr\!\left(X_{q(i)} \middle|\, \vect{y}^{(k-\ell)}\!\ldots \vect{y}^{(k-1)}\right) 
	},
\end{align*}
where the numerator represents the current belief of the \text{$k$-th} trellis about the input symbol $X_{q(i)}$, and the denominator corresponds to the belief previously received from the \text{$(k-1)$-th} trellis (at the $\ell$-th iteration). This approximation follows the same reasoning as in \eqref{equ:BPprior}. Similarly, the belief passed from the $k$-th trellis to the $(k+1)$-th trellis at the $(\ell+1)$-th iteration~is
\begin{align*}
	\Pr\!\big(X_{q(i)} \big|\, \vect{y}^{(k-\ell)} \!\ldots \vect{y}^{(k)} \big)\approx
	\frac{
		\Pr\!\big(X_{q(i)} \big|\, \vect{y}^{(k-\ell)}\!\ldots \vect{y}^{(k+\ell)}\big)
	}{
		\mathcal{Z} \cdot \Pr\!\left(X_{q(i)} \middle|\, \vect{y}^{(k+1)}\!\ldots \vect{y}^{(k+\ell)}\right)
	}.
\end{align*}

According to the outlined message-passing procedure, one can observe the propagation of extrinsic beliefs from local traces across the trellis network via the expansion of the observation window $\Psi_\ell^{(k)}$ after each iteration. It is straightforward to verify from \eqref{equ:BPprior} that after $K$ iterations the extrinsic prior
$
\Pr\bigl(X_{q(i)} \mid \vect{y}^{(1)},\ldots,\vect{y}^{(k-1)}, \vect{y}^{(k+1)},\ldots,\vect{y}^{(K)}\bigr)
$
is available at the $k$-th trellis for all $1 \le k \le K$. Applying the same steps as in \eqref{equ:rule}, \eqref{equ:finalapp}, and \eqref{equ:marginals:FB}, then yields the desired $\Lambda_i\big(X_{q(i)}\big)$.

By allowing a direct message-passing link between the first and the $K$-th trellis, simulation results further indicate that fewer than $K$ iterations suffice to achieve \emph{symbol-level consensus} for $K > 2$. The only exception is $K = 2$, for which fewer than five iterations suffice. Accordingly, assuming that the output pointer in each trace does not drift by more than $\Delta < N$ symbols from the input index and denoting the total number of state transitions by $\delta$, the complexity of computing APPs using the proposed method is upper-bounded by $\mathcal{O}\!\left(N \cdot \delta \Delta \cdot K^2\right)$. In contrast, MAP estimation via applying BCJR to the joint trellis has a complexity of $\mathcal{O}\!\left(N \cdot (\delta \Delta)^K\right)$.

\begin{algorithm}
	\caption{Proposed Belief-Combining Framework}\label{alg:BP}
	\textbf{Input:} $\pI, \pD, \pS, \vy^{(1)}, \ldots, \vy^{(K)}$\\[1pt]
	\For{$1 \leq k \leq K$}{
		\textit{Calculate} $\Pr\bigl(X_{q(i)}\bigm|\vy^{(k)}\bigr), \forall i,$ by marginalizing \eqref{equ:norm2};\!\!
	}
	$\ell \leftarrow 0$\;
	\While{$\neg \text{consensus}$}{
		$\ell \leftarrow \ell + 1$\;
		\For{$1 \leq k \leq K$}{
			\!\!\textit{Calculate} $\Pr\!\big(X_{q(i)} \,\big|\, \Psi_\ell^{(k)}\big), \forall i,$ from \eqref{equ:BPprior};\\[2pt]
			\!\!Using \eqref{equ:gamma:factor}, \eqref{equ:forward}, \eqref{equ:backward}, \eqref{equ:rule}, \eqref{equ:finalapp}, \textit{Calculate}:\\[1pt]
			\quad $\Pr\!\big(S_i^{(k)} | \vect{y}^{(k-\ell)}, \ldots, \vect{y}^{(k+\ell)}\big), \forall i$;\\[2pt]
			\!\!\textit{Calculate} $\Pr\!\big(X_{q(i)} \big| \vect{y}^{(k-\ell)}\!\!\!\ldots \vect{y}^{(k+\ell)}\big)\!, \forall i,$ from \eqref{equ:marginals:FB};\!\!\!\!\!\!\!\\[2pt]
			\!\!To use in \mbox{\tikz[baseline=-.5ex]{\node[fill=gray!15, draw=gray!50, line width=0.4pt, rounded corners=1.5pt, inner xsep=1pt, inner ysep=1pt] {\footnotesize\textsf{Step 9}};}}
			 of next iteration ($\ell\!+\!1$), \textit{Calculate}:\!\!\!\!\!\!\!\!\\[1pt]
			\quad $\Pr\!\big(X_{q(i)} \big|\, \vect{y}^{(k)}, \ldots, \vect{y}^{(k+\ell)}\big), \forall i$;\\[1pt]
			\quad $\Pr\!\big(X_{q(i)} \big|\, \vect{y}^{(k-\ell)}, \ldots, \vect{y}^{(k)} \big), \forall i$;
		}
	}
	\textbf{Return} $\Pr\!\big(X_t \big|\, \vy^{(1)}, \ldots, \vy^{(K)} \big), \forall t,$ from \mbox{\tikz[baseline=-.5ex]{\node[fill=gray!15, draw=gray!50, line width=0.33pt, rounded corners=1.5pt, inner xsep=1pt, inner ysep=1pt] {\footnotesize\textsf{Step 12}};}}
\end{algorithm}

The proposed framework is summarized in Algorithm~\ref{alg:BP}. Further, all associated Python codes are available online~\cite{IDS_Channels-BP_Trace_Reconstruction}.

\section{Simulation Results}\label{sec:results}
Since we consider the uncoded setting, we employ edit distance as our metric\textemdash because even a single edit can propagate Hamming errors across the entire sequence, precluding a fair evaluation of reconstruction capability under Hamming metric.

\begin{figure}
	\centering
	\includegraphics[scale=0.94]{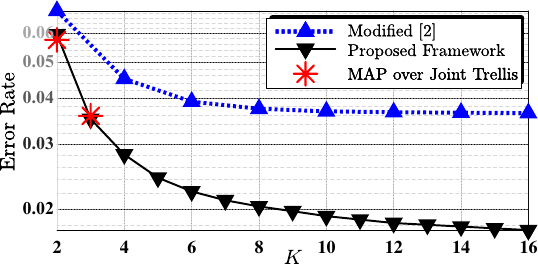}
	\caption{Error rates for uncoded real DNA reads ($N=110$).\vspace{-4pt}}\label{fig:sim_1}
\end{figure}

\begin{figure}
	\centering
	\includegraphics[scale=0.94]{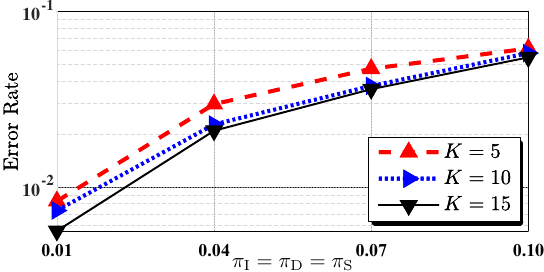}
	\caption{Error rates for uncoded random sequences ($N=100$).\vspace{4pt}}\label{fig:sim_2}
\end{figure}

In Fig.~\ref{fig:sim_1}, we evaluate the reconstruction performance of the proposed scheme across $1 < K \leq 16$ on a clustered dataset of nanopore sequencing reads with $N=110$, provided by~\cite{US2022166446A1}. The error rates for this dataset are $\pI=0.017$, $\pD=0.02$, and $\pS=0.022$. Each simulation point averages over $300$ randomly selected source sequences from a pool of $10{,}000$. Due to the exponential complexity of joint-trellis decoding, this baseline is used only for verification over $60$ sequences at $K=2$ and $10$ sequences at $K=3$. Unfortunately, we could not reproduce the uncoded results in~\cite{US2022166446A1}. Instead, we attempt to improve their approach by passing soft beliefs (rather than injecting hard estimates) from the decoding of each trace to subsequent traces, while maintaining the same complexity, $\mathcal{O}\!\left(N \cdot \delta \Delta \cdot K\right)$, as in~\cite{US2022166446A1}. We then evaluate its reconstruction performance on the same dataset using the edit metric (Fig.~\ref{fig:sim_1}). Ultimately, Fig.~\ref{fig:sim_2} illustrates the results for random sequences of length $N=100$ across various error rates ($\pI = \pD = \pS$).

\section{Conclusion}\label{sec:conclusion}
In this letter, we proposed an iterative belief-combining framework for multi-trace MAP estimation over IDS channels. Under a mild consensus condition, we proved that the resulting MAP estimates are identical to those derived from applying the BCJR algorithm to the joint multi-trace trellis, while decoding complexity scales only quadratically with the number of traces. Numerical simulations on both real DNA reads and randomly generated sequences across various error rates corroborated the analysis and consistently confirmed the predicted consensus.

As illustrated, even with $K=4$ traces, the proposed method achieves $97\%$ reconstruction fidelity on real DNA read data. Future work will extend this framework to a full-stack design, where simple inner codes are expected to suffice to achieve near-perfect symbol-level fidelity, thereby enabling simpler outer codes for strand disambiguation under Poisson sampling.

\bibliographystyle{IEEEtran}
\bibliography{citation_rep}

\end{document}